\documentclass[10pt]{amsart}
\usepackage{amssymb}
\usepackage{amsmath,amssymb,latexsym,amsfonts,amsthm,
latexsym, amscd, amsfonts, epsfig,epic,psfrag,eepic}
\usepackage{mathrsfs}
\usepackage{color}
\usepackage{eucal}%    caligraphic-euler fonts: \mathcal{ }
\usepackage{eufrak}%   frak-euler        fonts: \mathfrak{ }
\usepackage[all]{xypic}
\usepackage{xspace}
\usepackage{array}
\usepackage{graphicx}
\theoremstyle{plain}
\newtheorem{theorem}{Theorem}

\newtheorem{lemma}{Lemma}
\newtheorem{proposition}{Proposition}

\theoremstyle{definition}

\theoremstyle{example}

\theoremstyle{remark}

\numberwithin{equation}{section}

\begin{document}

\title[Irreducibility in RNA structures]
      {Irreducibility in RNA structures}
\author{Emma Y. Jin$^{\,\star}$ and Christian M. Reidys$^{\,\star,\dagger}$}
\address{$\star$ Center for Combinatorics, LPMC-TJKLC  \\
         $\dagger$ College of Life Sciences \\
         Nankai University  \\
         Tianjin 300071\\
         P.R.~China\\
         Phone: *86-22-2350-6800\\
         Fax:   *86-22-2350-9272}
\email{reidys@nankai.edu.cn}
\thanks{}
\date{January, 2009}
\begin{abstract}
In this paper we study irreducibility in RNA structures. By RNA structure
we mean RNA secondary as well as RNA pseudoknot structures. In
our analysis we shall contrast random and minimum free energy (mfe)
configurations. We compute various distributions: of the numbers of
irreducible substructures, their locations and sizes, parameterized
in terms of the maximal number of mutually crossing arcs, $k-1$,
and the minimal size of stacks $\sigma$. In particular, we analyze
the size of the largest irreducible substructure for random and mfe
structures, which is the key factor for the folding time of mfe
configurations.
\end{abstract}
\maketitle {{\small
%\tableofcontents
}}
%%%
%%%
%%%%%%%%%%%%%%%%%%%%%%%%%%%%%%%%%%%%%%%%%%%%%%%%%%%%%%%%%%%%%%%%%%%%%%%%
%%%
%%%
\section{Introduction and background}

%%%
%%%%%%%%%%%%%%%%%%%%%%%%%%%%%%%%%%%%%%%%%%%%%%%%%%%%%%%%%%%%%%%%%%%%%%%%
%%%

In this paper we study irreducibility in RNA structures.
Intuitively, an irreducible substructure over a subsequence is a
configuration of bonds, beginning and ending with arcs of certain
stack size, that cannot be written as a nontrivial concatenation of
smaller configurations. Since any minimum free energy (mfe) folding
algorithm depends at least polynomially (to a degree larger than
one) on the sequence length, the size of the largest, irreducible
substructure determines the folding time.

\begin{figure}[ht]
\centerline{%
\epsfig{file=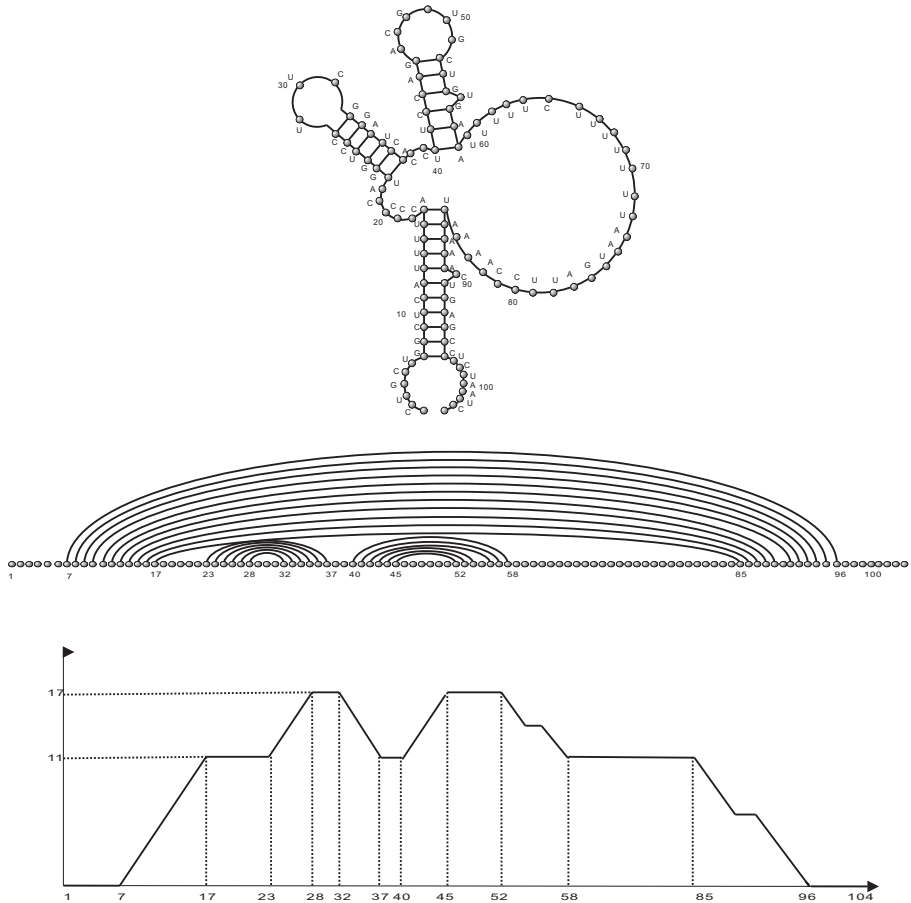,width=0.80\textwidth}\hskip15pt }
\caption{\small The RNA secondary structure of the 3'-UTR of subnuclei mRNA,
represented as planar graph, diagram and Motzkin-path.}
\label{F:sec}
\end{figure}

Let us begin by recalling some basic facts about RNA structures: an
RNA structure is the helical configuration of its primary sequence,
i.e.~the sequence of nucleotides {\bf A}, {\bf G}, {\bf U} and {\bf
C}, together with Watson-Crick ({\bf A-U}, {\bf G-C}) and ({\bf
U-G}) base pairs. One well-known class of RNA structures, are RNA
secondary structures, pioneered three decades ago by Waterman
\cite{Penner:93c,Waterman:79a,Waterman:78a,Waterman:80,Waterman:94a}.
Secondary structures exhibit exclusively noncrossing bonds and are
subject to specific minimum arc-length conditions. They can readily
be identified with Motzkin-paths satisfying some minimum height and
plateau-length, see Figure~\ref{F:sec} \cite{Waterman:94a}.
%%%
%%%%%%%%%%%%%%%%%%%%%%%%%%%%%%%%%%%%%%%%%%%%%%%%%%%%%%%%%%%%%%%%%%%%%%%%
%%%
%%%
%%%%%%%%%%%%%%%%%%%%%%%%%%%%%%%%%%%%%%%%%%%%%%%%%%%%%%%%%%%%%%%%%%%%%%%%
%%%
The latter restrictions come from biophysical constraints due to mfe
loop-energy parameters and limited flexibility of bonds. It is clear
from the above bijection, that irreducible substructures in RNA
secondary structures are closely related to the number of nontrivial
returns, i.e.~the number of non-endpoints, for which the
Motzkin-path meets the $x$-axis. As a purely combinatorial problem
this has been studied by \cite{Cameron:03,Reidys:09hit}.

\begin{figure}[ht]
\centerline{%
\epsfig{file=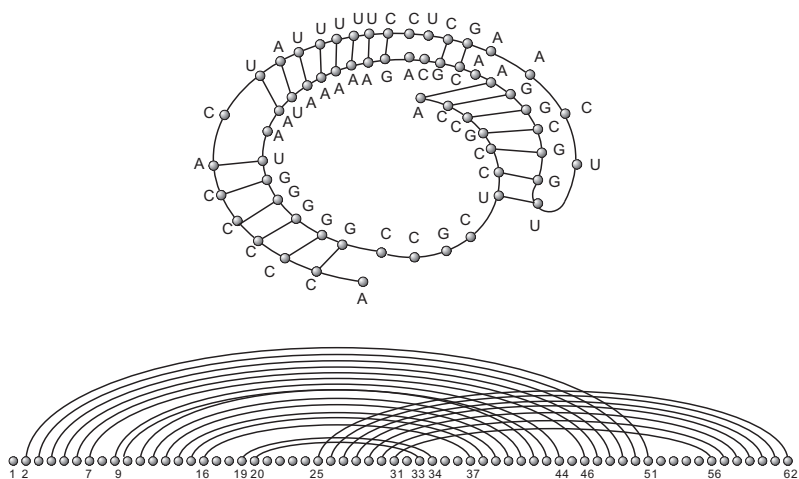,width=0.7\textwidth}\hskip15pt
 }
\caption{\small The mRNA plasmid pMU$720$ (IncB)-pseudoknot
structure: its planar graph (top) and diagram representation
(bottom).} \label{F:plasmid.eps}
\end{figure}

It is well-known that RNA configurations are far more complex than
secondary structures: they exhibit additional, cross-serial
nucleotide interactions \cite{Searls}. These interactions were
observed in natural RNA structures, as well as via comparative
sequence analysis \cite{Westhof:92a}. They are called pseudoknots,
see Figure~\ref{F:plasmid.eps}, and widely occur in functional RNA,
like for instance, eP RNA \cite{Loria:96a} as well as ribosomal RNA
\cite{Konings:95a}. RNA pseudoknots are conserved also in the
catalytic core of group I introns. In plant viral RNAs pseudoknots
mimic tRNA structure and in vitro RNA evolution \cite{Tuerk:92}
experiments have produced families of RNA structures with pseudoknot
motifs, when binding HIV-1 reverse transcriptase.
%%%
%%%%%%%%%%%%%%%%%%%%%%%%%%%%%%%%%%%%%%%%%%%%%%%%%%%%%%%%%%%%%%%%%%%%%%%%%%%
%%%
%%%
%%%%%%%%%%%%%%%%%%%%%%%%%%%%%%%%%%%%%%%%%%%%%%%%%%%%%%%%%%%%%%%%%%%%%%%%%%%
%%%

\begin{figure}[ht]
\centerline{%
\epsfig{file=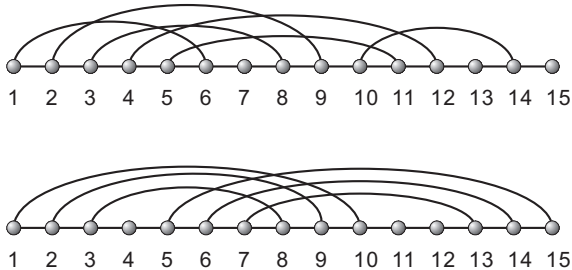,width=0.5\textwidth}\hskip15pt
 }
\caption{\small $k$-(nc) diagrams: we display a $4$-(nc), arc-length
$\lambda\ge 4$ and $\sigma\ge 1$ diagram (top), where the edge set
$\{(1,6),(2,9),(4,12)\}$ is a $3$-crossing, the arc $(10,14)$ has
length $4$ and $(1,6)$ has stack-length $1$. Below, we display a
$3$-(nc), $\lambda\ge 5$ and $\sigma\ge 3$ (lower) diagram, where
$(3,8)$ has arc-length $5$ and the stack $((1,10),(2,9),(3,8))$ has
stack-length $3$.} \label{F:dia}
\end{figure}

\begin{figure}[ht]
\centerline{%
\epsfig{file=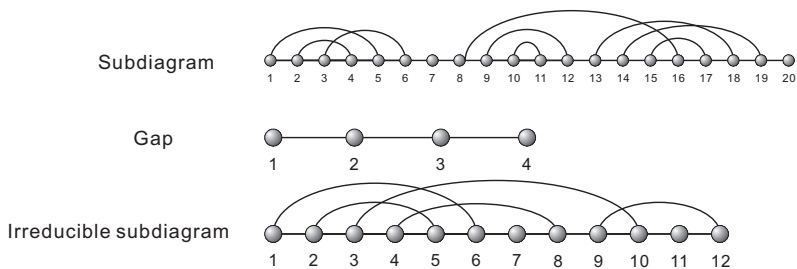,width=0.7\textwidth}\hskip15pt
 }
\caption{\small Subdiagrams, gaps and irreducibility: a diagram
(top), decomposed into the subdiagram over $(1,6)$, the gap $7$ and
the subdiagram over $(8,19)$ and gap $20$. A gap (middle) and an
irreducible diagram over $(1,12)$ (bottom).} \label{F:gapsub}
\end{figure}

\begin{figure}[ht]
\centerline{%
\epsfig{file=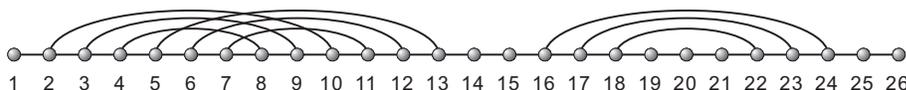,width=0.8\textwidth}\hskip15pt
 }
\caption{\small A $3$-(nc), $3$-(ca) RNA structure has returns at
position $13$ and $24$, respectively.} \label{F:defreturn}
\end{figure}

\begin{figure}[ht]
\centerline{%
\epsfig{file=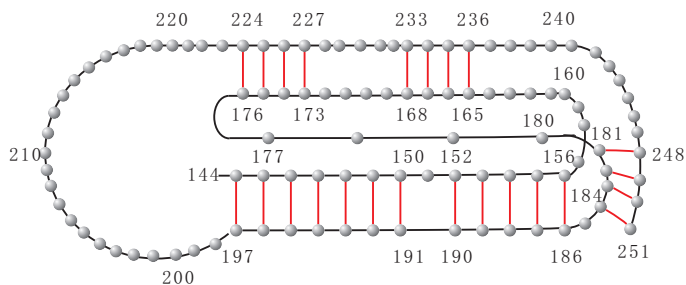,width=0.6\textwidth}\hskip15pt
 }
\caption{\small mRNA-Ec$_\alpha$: the irreducible pseudoknot
structure of the regulatory region of the $\alpha$ ribosomal protein
operon.} \label{F:pseudo1}
\end{figure}

\begin{figure}[ht]
\centerline{%
\epsfig{file=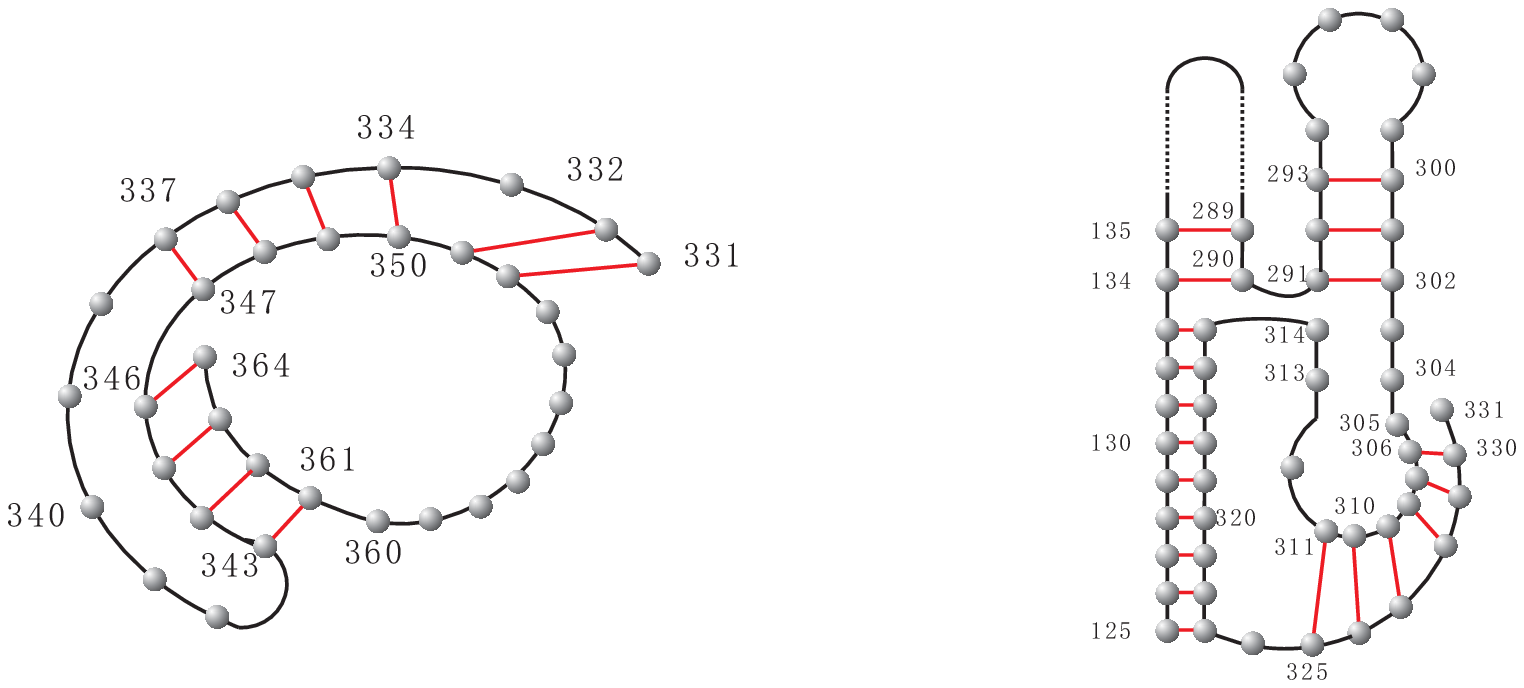,width=0.6\textwidth}\hskip15pt
 }
\caption{\small Hepatitis C Virus (left) and Hepatitis C Virus-IRES
(right): the largest irreducible substructure is of size almost
$n$.} \label{F:HCV}
\end{figure}

Combinatorially, cross serial interactions are tantamount to crossing bonds.
Therefore, RNA pseudoknot structures have been modeled as $k$-noncrossing
($k$-(nc))
diagrams \cite{Reidys:07pseu,Reidys:07lego,Reidys:08ma},
i.e.~labeled graphs over the vertex set $[n]=\{1,\dots, n\}$ with degree
$\le 1$. Diagrams are represented by drawing their vertices $1,\dots,n$
in a horizontal line and their arcs $(i,j)$, where $i<j$, in the upper
half-plane. Here the degree of $i$ refers to the number of non-horizontal arcs
incident to $i$, i.e.~the backbone of the primary sequence is not
considered. The vertices and arcs correspond to nucleotides and
Watson-Crick ({\bf A-U}, {\bf G-C}) and ({\bf U-G}) base pairs,
respectively, see Figure~\ref{F:dia}.
%%%
%%%%%%%%%%%%%%%%%%%%%%%%%%%%%%%%%%%%%%%%%%%%%%%%%%%%%%%%%%%%%%%%%%%%%%%%%%%
%%%
%%%
%%%%%%%%%%%%%%%%%%%%%%%%%%%%%%%%%%%%%%%%%%%%%%%%%%%%%%%%%%%%%%%%%%%%%%%%%%%
%%%
Diagrams are characterized via their maximum number of mutually
crossing arcs, $k-1$, their  minimum arc-length, $\lambda$, and their
minimum stack-length, $\sigma$.
A $k$-crossing is a set of $k$ distinct arcs
$(i_{1},j_{1}),(i_{2},j_{2}),\ldots(i_{k},j_{k})$
with the property $i_{1}<i_{2}<\ldots<i_{k}<j_{1}<j_{2}<\ldots<j_{k}$. A
diagram without any $k$-crossings is called a $k$-(nc) diagram.
The length of an arc $(i,j)$ is $j-i$ and a stack of length $\sigma$
is a sequence of ``parallel'' arcs of the form
\begin{equation*}
((i,j),(i+1,j-1),\dots,(i+(\sigma-1),j-(\sigma-1))).
\end{equation*}
A subdiagram of a $k$-(nc) diagram is a subgraph over a
subset $M\subset [n]$ of consecutive vertices that starts with an
origin and ends with a terminus of some arc.
Let $(i_1,\dots,i_m)$ be a sequence of isolated points, and
$(j_1,j_2)$ be an arc. We call $(i_1,\dots,i_m)$ interior if and
only if there exists some arc $(j_1,j_2)$ such that
$j_1<i_1<i_m<j_2$ holds and exterior, otherwise.
Any exterior sequence of consecutive, isolated vertices is called a gap.
A diagram and subdiagram is called irreducible, if it cannot be decomposed
into a (nontrivial) sequence of gaps and subdiagrams, see
Figure~\ref{F:gapsub}. As a result, any $k$-(nc) diagram can
be uniquely decomposed into an alternating sequence of gaps and
irreducible subdiagrams.
%%%%%%%%%%%%%%%
%%%%%%%%%%%%%%%%
We call a $k$-(nc), $\sigma$-canonical ($\sigma$-(ca)) diagram
with arc-length $\ge 4$
and stack-length $\ge \sigma$, a $k$-(nc), $\sigma$-(ca) RNA
structure, see Figure~\ref{F:dia}. We accordingly adopt the notions
of gap, substructure and irreducibility for RNA structures. A
$k$-(nc), $\sigma$-(ca) RNA structure has return at position $i$ if
$i$ is the endpoint of some irreducible substructure, see
Figure~\ref{F:defreturn}.
%%%
%%%%%%%%%%%%%%%%%%%%%%%%%%%%%%%%%%%%%%%%%%%%%%%%%%%%%%%%%%%%%%%%%%%%%%
%%%
%%%
%%%%%%%%%%%%%%%%%%%%%%%%%%%%%%%%%%%%%%%%%%%%%%%%%%%%%%%%%%%%%%%%%%%%%%
%%%
Unique large irreducible substructures are quite common for natural
RNA pseudoknot structures, see Figure~\ref{F:pseudo1}. The size of the
largest irreducible substructure is typically very large: it contains
almost all nucleotides, see Figure~\ref{F:HCV}.
%%%
%%%%%%%%%%%%%%%%%%%%%%%%%%%%%%%%%%%%%%%%%%%%%%%%%%%%%%%%%%%%%%%%%%%%%%
%%%
%%%
%%%%%%%%%%%%%%%%%%%%%%%%%%%%%%%%%%%%%%%%%%%%%%%%%%%%%%%%%%%%%%%%%%%%%%
%%%
%%%
%%%%%%%%%%%%%%%%%%%%%%%%%%%%%%%%%%%%%%%%%%%%%%%%%%%%%%%%%%%%%%%%%%%%%%
%%%
%%%
%%%%%%%%%%%%%%%%%%%%%%%%%%%%%%%%%%%%%%%%%%%%%%%%%%%%%%%%%%%%%%%%%%%%%%%%%%
%%%

The paper is organized as follows: in Section~\ref{S:theory} we
recall some combinatorial framework due to \cite{Reidys:09hit}. In
particular, we derive the probability generating function for
the number of irreducible substructures. We remark that the
framework presented in Section~\ref{S:theory} can be generalized to
RNA tertiary structures. In Section~\ref{S:comp1} we put these
results to the test: we shall compare random and mfe structures. We
begin by observing specific deviations of the distributions of mfe
$2$- and $3$-(nc) structures for $n=75$ and $\sigma=3$ from that of
random structures. The rest of the section we analyze these
deviations and prove in the process (Proposition~\ref{P:elong}) a
``shift''-result. The latter allows us to understand the effect of
increasing the stack-size $\sigma$ on irreducibility.
In Section~\ref{S:return} we derive simple
formulas to the probabilities of return locations, i.e.~the
endpoints of irreducible substructures and contrast random and mfe
secondary and pseudoknot structures. In Section~\ref{S:sizecomp} we
study the size of the largest irreducible components for $k$-(nc),
$\sigma$-(ca) RNA structures.

%%%
%%%%%%%%%%%%%%%%%%%%%%%%%%%%%%%%%%%%%%%%%%%%%%%%%%%%%%%%%%%%%%%%%%%%%%
%%%
\section{Some combinatorics}\label{S:theory}
%%%
%%%%%%%%%%%%%%%%%%%%%%%%%%%%%%%%%%%%%%%%%%%%%%%%%%%%%%%%%%%%%%%%%%%%%%
%%%
Let $\delta_{n,j}^{(k,\sigma)}$ denote the number of $k$-(nc),
$\sigma$-(ca) RNA structures, containing exactly $j$ irreducible
substructures and let $\delta_{n}^{(k,\sigma)}=\sum_{j\ge
0}\delta_{n,j} ^{(k,\sigma)}$. That is, $\delta_{n}^{(k,\sigma)}$
denotes the number of $k$-(nc), $\sigma$-(ca) RNA structures. The
bivariate generating function of the $\delta_{n,j}^{(k,\sigma)}$
indexed by $j$, the number of irreducible substructures and $n$, the
sequence length is given by
\begin{equation}\label{E:JJ}
{\bf U}_{k,\sigma}(z,u)=\sum_{n\ge 0}\sum_{j\ge
0}\delta_{n,j}^{(k,\sigma)}u^jz^n.
\end{equation}
Let furthermore ${\bf T}_{k,\sigma}(z)= \sum_{n\ge 0}\delta_{n}^{(k,\sigma)}
z^n$ and ${\bf R}_{k,\sigma}(z)$ denote the generating function of irreducible
RNA structures. The following lemma \cite{Reidys:09hit} derives the generating
function ${\bf U}_{k,\sigma}(z,u)$:
%%%
%%%%%%%%%%%%%%%%%%%%%%%%%%%%%%%%%%%%%%%%%%%%%%%%%%%%%%%%%%%%%%%%%%%%%%%%%%%%%
%%%
\begin{lemma}\label{L:bistar}
The bivariate generating function of the number of $k$-(nc),
$\sigma$-(ca) RNA structures, which contain exactly $j$ irreducible
$k$-(nc), $\sigma$-(ca) RNA substructures, is given by
\begin{eqnarray*}
{\bf U}_{k,\sigma}(z,u) &=&\frac{\frac{1}{1-z}}
{1-u\left(1-\frac{1}{(1-z){\bf T}_{k,\sigma}(z)}\right)}.\\
\end{eqnarray*}
\end{lemma}
%%%
%%%%%%%%%%%%%%%%%%%%%%%%%%%%%%%%%%%%%%%%%%%%%%%%%%%%%%%%%%%%%%%%%%%%%%%%%%%%%
%%%
Lemma~\ref{L:bistar} is the key for computing the limit distribution
of the number of $k$-(nc), $\sigma$-(ca) RNA structures that have
exactly $j$ irreducible RNA substructures. For this purpose, let
$\xi_n^{(k,\sigma)}$ be the r.v. having the probability distribution
\begin{eqnarray}
\mathbb{P}(\xi_n^{(k,\sigma)}=j)=
\frac{\delta_{n,j}^{(k,\sigma)}}{\delta_{n}^{(k,\sigma)}}.
\end{eqnarray}
The theorem below \cite{Reidys:09hit} shows that the probabilities
$\mathbb{P}(\xi_n^{(k,\sigma)}=j)$ satisfy a discrete limit law:
%%%
%%%%%%%%%%%%%%%%%%%%%%%%%%%%%%%%%%%%%%%%%%%%%%%%%%%%%%%%%%%%%%%%%%%%%%%%%%%%%
%%%
\begin{theorem}{\label{T:centra2}}
Let $\alpha_{k,\sigma}$ be the real positive dominant singularity of
${\bf T}_{k,\sigma}(z)$ and
$$\tau_{k,\sigma}=1-\frac{1}{(1-\alpha_{k,\sigma}){\bf
T}_{k,\sigma}(\alpha_{k,\sigma})}.$$ Then the
r.v.~$\xi_n^{(k,\sigma)}$ satisfies the discrete limit law
\begin{equation}
\lim_{n\rightarrow
\infty}\mathbb{P}(\xi_{n}^{(k,\sigma)}=i)=
\frac{(1-\tau_{k,\sigma})^2}{\tau_{k,\sigma}}\,i
\tau_{k,\sigma}^{i}.
\end{equation}
That is, $\xi_n^{(k,\sigma)}$ is determined by the density function
of a $\Gamma(-\ln\tau_{k,\sigma},2)$-distribution. Furthermore, the
probability generating function probability generating function of
the limit distribution is given by
\begin{eqnarray*}
q(u)=\sum_{n\ge 1}\mathbb{P}(\xi_{n}^{(k,\sigma)}=i)u^i
=\frac{u(1-\tau_{k,\sigma})^2} {(1-\tau_{k,\sigma} u)^2}.
\end{eqnarray*}
\end{theorem}
%%%
%%%%%%%%%%%%%%%%%%%%%%%%%%%%%%%%%%%%%%%%%%%%%%%%%%%%%%%%%%%%%%%%%%%%%%%%%%%%%
%%%
The generating function of $k$-(nc), $\sigma$-(ca) RNA structures,
${\bf T}_{k,\sigma}(z)$ and its dominant singularities,
$\alpha_{k,\sigma}$, have been studied in
\cite{Reidys:07pseu,Reidys:07lego,Reidys:08ma}. In particular the
limiting probability of irreducible RNA structures is given by
\begin{equation}
\lim_{n\rightarrow
\infty}\mathbb{P}(\xi_{n}^{(k,\sigma)}=1)=(1-\tau_{k,\sigma})^2.
\end{equation}
We observe that for fixed $\sigma$ and increasing crossing number, $k$, the
singularity $\tau_{k,\sigma}$ decreases. Therefore the limiting probability
of RNA structures to be irreducible increases with increasing crossing number.
However, for fixed $k$ and increasing $\sigma$, the singularity $\tau_{k,
\sigma}$ increases. Consequently, the limiting probability of RNA
structures to be irreducible decreases with increasing $\sigma$.
Theorem~\ref{T:centra2} allows to compute the characteristic
function of the r.v. $\xi_{n}^{(k,\sigma)}$:
\begin{equation*}
\mathbb{E}[e^{it\xi_{n}^{(k,\sigma)}}] = q(e^{it})
=\frac{e^{it}(1-\tau_{k,\sigma})^2} {(1-\tau_{k,\sigma} e^{it})^2}.
\end{equation*}
By Taylor expansion of the characteristic function we obtain the
$k$-th moments of $\xi_{n}^{(k,\sigma)}$, i.e.,
\begin{eqnarray}
\mathbb{E}[e^{it\xi_{n}^{(k,\sigma)}}]=
1+(it)\mathbb{E}[\xi_{n}^{(k,\sigma)}]+
\frac{(it)^2}{2!}\mathbb{E}[(\xi_{n}^{(k,\sigma)})^2]
+\cdots+\frac{(it)^m}{m!}\mathbb{E}[(\xi_{n}^{(k,\sigma)})^m]+o(t).
\end{eqnarray}
Consequently, we can compute expectation and variance of $\xi_{n}^{(k,\sigma)}$
for varying $k$ and $\sigma$, see Table~\ref{Ta:twin1} and
Table~\ref{Ta:twin2}:
\begin{table}[htbp]
\begin{minipage}[t]{2.65in}
\centering
\parbox[t]{2.30in}{\caption{$k=2$}\label{Ta:twin1}}\\
\begin{tabular}{c|cccccccccccccccccc}
& $\tau_{2,\sigma}$ &$\mathbb{E}[\xi_{n}^{(2,\sigma)}]$ &
$\mathbb{V}[\xi_{n}^{(2,\sigma)}]$ \\
\hline $\sigma=3$ & \small{0.3201} & \small{1.9416} & \small{5.1548}\\
\hline $\sigma=4$ & \small{0.3441} & \small{2.0492} &\small{5.7991}\\
\hline $\sigma=5$ & \small{0.3615} & \small{2.1323} &\small{6.3203}\\
\hline
\end{tabular}
\end{minipage}
\hfill
\begin{minipage}[t]{2.65in}
\parbox[t]{2.30in}{\caption{$k=3$}\label{Ta:twin2}}\\
\centering
\begin{tabular}{c|cccccccccccccccccc}
& $\tau_{3,\sigma}$ & $\mathbb{E}[\xi_{n}^{(3,\sigma)}]$ &
$\mathbb{V}[\xi_{n}^{(3,\sigma)}]$ \\
\hline $\sigma=3$ &\small{0.0167} & \small{1.0340} & \small{1.1036}\\
\hline $\sigma=4$ &\small{0.0208} & \small{1.0425} &\small{1.1302}\\
\hline $\sigma=5$ &\small{0.0244} & \small{1.0500} &\small{1.1538}\\
\hline
\end{tabular}
\end{minipage}
\end{table}
Table~\ref{Ta:twin1} shows, that for RNA secondary structures
increasing the stack size $\sigma$ significantly increases
reducibility. Table~\ref{Ta:twin2} indicates that $3$-(nc) RNA
pseudoknot structures are typically irreducible with rather subtle
dependence on the minimum stack-size, $\sigma$.

%%%
%%%%%%%%%%%%%%%%%%%%%%%%%%%%%%%%%%%%%%%%%%%%%%%%%%%%%%%%%%%%%%%%%%%%%%%%%
%%%

\section{RNA random structures and RNA mfe structures}\label{S:comp1}

%%%
%%%%%%%%%%%%%%%%%%%%%%%%%%%%%%%%%%%%%%%%%%%%%%%%%%%%%%%%%%%%%%%%%%%%%%%%%
%%%

\begin{figure}[ht]
\centerline{%
\epsfig{file=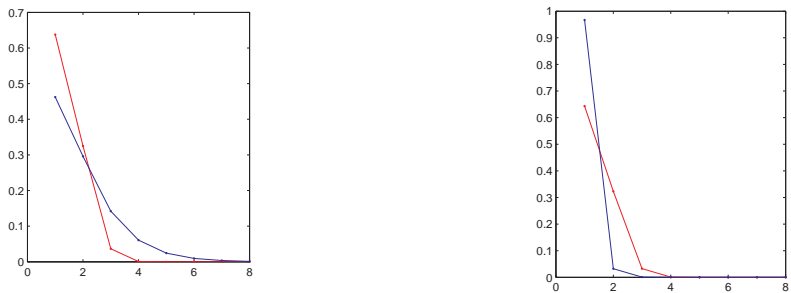,width=0.9\textwidth}\hskip15pt
 }
\caption{\small Random versus mfe: the lhs shows the distribution of
irreducibles in 2-(nc), 3-(ca) mfe (red) and random structures
(blue), for $n=75$. The rhs showcases these distributions for
3-(nc), 3-(ca) mfe (red) and random structures (blue) for $n=75$.}
\label{F:experiment}
\end{figure}

In this section we analyze irreducibility in random and mfe RNA
secondary and pseudoknot-structures. As folding algorithms for the
generation of the mfe RNA secondary and pseudoknot structures we
employ Vienna RNA \cite{Vienna} and {\sf cross}
\cite{Reidys:08algo}. We shall begin by comparing in
Figure~\ref{F:experiment} irreducibility of $2$-(nc) and $3$-(nc)
random and mfe structures (of length $n=75$) for minimum stack size
$\sigma=3$. Figure~\ref{F:experiment} shows that: (a) for $2$-(nc)
$3$-(ca) structures the mfe structures are more irreducible than
their random counterparts and (b) for $3$-(nc) structures the
contrary is being observed: $3$-(nc) mfe structures are less
irreducible than $3$-(nc) random structures.
%%%
%%%%%%%%%%%%%%%%%%%%%%%%%%%%%%%%%%%%%%%%%%%%%%%%%%%%%%%%%%%%%%%%%%%%%%%%%%
%%%
%%%
%%%%%%%%%%%%%%%%%%%%%%%%%%%%%%%%%%%%%%%%%%%%%%%%%%%%%%%%%%%%%%%%%%%%%%%%%%
%%%

In order to understand the above observations, let us proceed by
analyzing first the effect of increasing the minimum stack-size
$\sigma$ for $2$- and $3$-(nc) mfe structures. While we have shown
in Section~\ref{S:theory} that the limit distribution shifts towards
{\it less} irreducibility when increasing $\sigma$,
Figure~\ref{F:23-33-75-85} shows that for fixed $\sigma$, $2$-(nc)
as well as $3$-(nc) mfe structures become {\it less} irreducible
when the sequence length $n$ increases. Intuitively, the increase in
$\sigma$ for fixed $n$ implies that any irreducible substructures
has to become larger. Therefore, in light of the fact that there are
only a few irreducible substructures, disallowing for these small
irreducible substructures implies the shift towards irreducibility.
With this picture in mind, we shall proceed by quantifying this
phenomenon:
%%%
%%%%%%%%%%%%%%%%%%%%%%%%%%%%%%%%%%%%%%%%%%%%%%%%%%%%%%%%%%%%%%%%%%%%%%%%%%%%
%%%
\begin{figure}[ht]
\centerline{
\epsfig{file=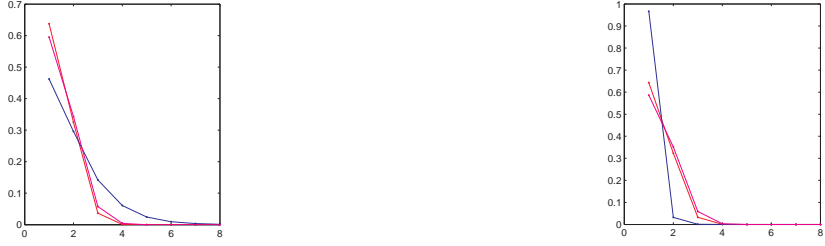,width=0.9\textwidth}\hskip15pt
 }
\caption{\small Finite size effect: here we insert the distribution
of irreducibles of 2-(nc), 3-ca mfe structures for $n=85$ (magenta)
into Figure~\ref{F:experiment}. }\label{F:23-33-75-85}
\end{figure}
%%%
%%%%%%%%%%%%%%%%%%%%%%%%%%%%%%%%%%%%%%%%%%%%%%%%%%%%%%%%%%%%%%%%%%%%%%%%%%%%
%%%
let
\begin{equation*}
R_{k,\sigma}(n)=[z^n]{\bf R}_{k,\sigma}(z)=[z^n]\left[-\frac{1}{{\bf
T}_{k,\sigma}(z)}\right],
\end{equation*}
i.e.~$R_{k,\sigma}(n)$ denotes the number of irreducible structures
over $n$ nucleotides and $\beta_{k,\sigma,j}(n)$ denotes the number
of $k$-(nc), $\sigma$-(ca) RNA structures with exactly $j$
irreducible substructures. Then, clearly,
$R_{k,\sigma}(n)<\beta_{k,\sigma,1}(n)$. We shall prove that, for
sufficiently large $n$, the scaling factor needed for passing from
$\sigma$ to $\sigma+1$ is
$\frac{\ln(\alpha_{k,\sigma})}{\ln(\alpha_{k,\sigma+1})}$.
%%%
%%%%%%%%%%%%%%%%%%%%%%%%%%%%%%%%%%%%%%%%%%%%%%%%%%%%%%%%%%%%%%%%%%%%%%%%%%%%
%%%
\begin{proposition}\label{P:elong}
For sufficiently large $n$ and arbitrary $\sigma$, we have
\begin{eqnarray}{\label{E:elong}}
R_{k,\sigma}(n)\sim \left(\frac{\ln(\alpha_{k,\sigma})}
{\ln(\alpha_{k,\sigma+1})}\right)^{-\mu}
R_{k,\sigma+1}\left(\left[\frac{\ln(\alpha_{k,\sigma})}
{\ln(\alpha_{k,\sigma+1})}\right]n\right).
\end{eqnarray}
Furthermore, $\beta_{k,\sigma,j}(n)$ satisfies
\begin{eqnarray}{\label{E:elongirre}}
\beta_{k,\sigma+1,j}\left[\frac{\ln(\alpha_{k,\sigma})}
{\ln(\alpha_{k,\sigma+1})}\cdot
n\right]\approx\beta_{k,\sigma,j}(n).
\end{eqnarray}
\end{proposition}
%%%
%%%%%%%%%%%%%%%%%%%%%%%%%%%%%%%%%%%%%%%%%%%%%%%%%%%%%%%%%%%%%%%%%%%%%%%%%%%%
%%%

In order to illustrate Proposition~\ref{P:elong}, we consider
$2$-(nc), $3$-(ca) and $2$-(nc), $4$-(ca) structures. According to
eq.~(\ref{E:gut})
\begin{equation*}
x^{(2)}_{75}=\left\lfloor
n\left(\frac{\ln(\alpha_{k,\sigma})}{\ln(\alpha_{k,\sigma}+1)}-1\right)
\right\rfloor\bigg|_{n=75,k=2,\sigma=3}=\left\lfloor
75\left(\frac{\ln(0.6053)}{\ln(0.6504)}-1\right)\right\rfloor=12
\end{equation*}
and for $3$-(nc), $3$-(ca) and $3$-(nc), $4$-(ca) structures we
obtain
\begin{equation*}
x_{75}^{(3)}=\left\lfloor
n\left(\frac{\ln(\alpha_{k,\sigma})}{\ln(\alpha_{k,\sigma}+1)}-1\right)
\right\rfloor\bigg|_{n=75,k=3,\sigma=3}=\left\lfloor
75\left(\frac{\ln(0.4914)}{\ln(0.5587)}-1\right)\right\rfloor =16.
\end{equation*}
%%%%
%%%%%%%%%%%%%%%%%%%%%%%%%%%%%%%%%%%%%%%%%%%%%%%%%%%%%%%%%%%%%%%%%%%%%%%%%%%%%%
%%%
\begin{figure}[ht]
\centerline{%
\epsfig{file=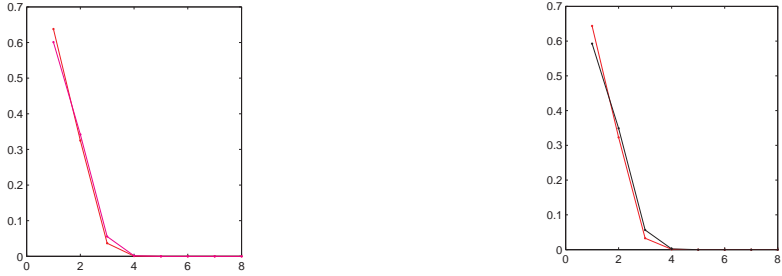,width=0.9\textwidth}\hskip15pt
 }
\caption{\small Proposition~\ref{P:elong} at work: the lhs shows
$2$-(nc), $3$-(ca) mfe structures for $n=75$ (red) and $2$-(nc),
$4$-(ca) mfe structures for $n=85$ (magenta). The rhs displays
$3$-(nc), $3$-(ca) mfe structures for $n=75$ (red) and $3$-(nc)
$4$-(ca) mfe structures for $n=90$ (black).}\label{F:shift}
\end{figure}
%%%%
%%%%%%%%%%%%%%%%%%%%%%%%%%%%%%%%%%%%%%%%%%%%%%%%%%%%%%%%%%%%%%%%%%%%%%%%%%%%%%
%%%
Figure~\ref{F:shift} shows how well the ``shifting'' works for $2$-
and $3$-(nc), $3$-(ca) mfe RNA structures.
%%%%
%%%%%%%%%%%%%%%%%%%%%%%%%%%%%%%%%%%%%%%%%%%%%%%%%%%%%%%%%%%%%%%%%%%%%%%%%%%%%%
%%%
\begin{figure}[ht]
\centerline{%
\epsfig{file=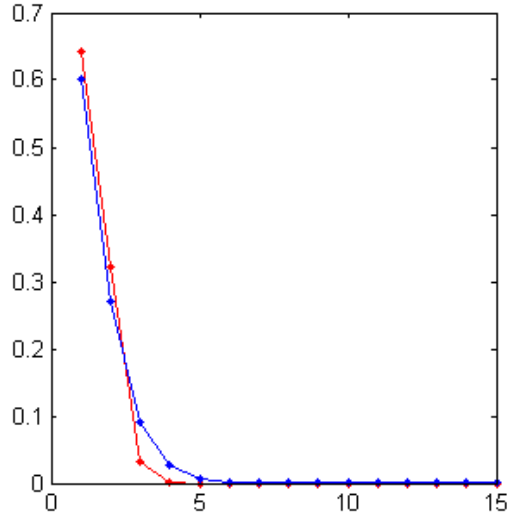,width=0.45\textwidth}\hskip15pt
 }
\caption{\small The distributions of irreducibles of $2$-(nc),
$1$-(ca) mfe (red) and random structures (blue) for $n=75$.
}\label{F:match}
\end{figure}
%%%%
%%%%%%%%%%%%%%%%%%%%%%%%%%%%%%%%%%%%%%%%%%%%%%%%%%%%%%%%%%%%%%%%%%%%%%%%%%%%%%
%%%
Proposition~\ref{P:elong} brings us now in the position to
understand observation (a): the discrepancy of the irreducibility of
$2$-(nc), $3$-(ca) mfe structures of length $75$ and random
structures. To this end we compare in Figure~\ref{F:match} the
irreducibility of $2$-(nc), $1$-(ca) mfe structures \cite{Vienna}
for $n=75$ with that of random structures. In view of
\begin{equation*}
\Delta_{75}= \left\lfloor
75\left(\frac{\ln(\alpha_{2,1})}{\ln(\alpha_{2,3})}-1\right)\right\rfloor
\approx \left\lfloor 75\left(
\frac{\ln(0.4369)}{\ln(0.6053)}-1\right)\right\rfloor=48,
\end{equation*}
the former correspond to $2$-(nc), $3$-(ca) structures of length
$75+48=123$. Accordingly, Proposition~\ref{P:elong} and
Figure~\ref{F:match} imply that $2$-(nc), $3$-(ca) mfe structures of
length $123$ exhibit an almost identical distribution as random
structures of infinite sequence length. Finally, we remark upon the
``paradox'' that Proposition~\ref{P:elong}--being entirely based on
the limit distribution of random structures--allows us to obtain
information about mfe structures of length $75$.

\begin{figure}[ht]
\centerline{%
\epsfig{file=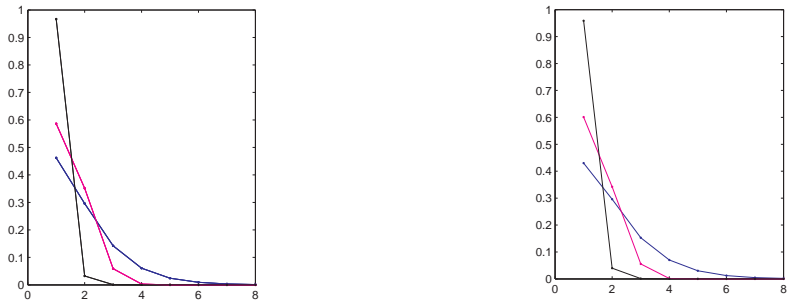,width=0.9\textwidth}\hskip15pt
 }
\caption{\small $2$- and $3$-(nc) random structures (blue/black)
versus $3$-(nc) mfe structures (magenta) for $n=85$. The lhs and rhs
display these curves for $\sigma=3$ and $\sigma=4$,
respectively.}\label{F:33-34-data}
\end{figure}

As for observation (b), recall that $3$-(nc) structures consist of
two distinct combinatorial classes: $2$-(nc) and $2$-crossing RNA
structures. The combinatorics of $2$- and $3$-(nc) RNA structures
\cite{Reidys:08ma}, implies that the number of $2$-crossings is
exponentially larger than the number of $2$-(nc) RNA structures, see
Table~\ref{T:1}.
\begin{table}
\begin{center}
\begin{tabular}{|c|c|c|c|c|c|c|c|c|c|}
%\hline
%  \multicolumn{9}{|c|}{\textbf{$\lambda=2$}}\\
\hline $k$ & \small$2$ & \small$3$ & \small $4$ & \small $5$ & \small $6$
&\small $7$ & \small $8$ & \small $9$  \\
\hline $\sigma=3$ &\small $1.6521$& \small$2.0348$ & \small $2.2644$ & \small
$2.4432$ & \small $2.5932$
&\small $2.7243$ & \small $2.8414$ & \small $2.9480$  \\
$\sigma=4$ & \small$1.5375$& \small$1.7898$ & \small $1.9370$ &
\small $2.0488$ & \small $2.1407$ &\small $2.2198 $ & \small $2.2896$ &
\small $2.3523$  \\
$\sigma=5$ & \small $1.4613$ & \small$1.6465$ & \small $1.7532$ &
\small $1.8330$ & \small $1.8979$
&\small $1.9532$ & \small $2.0016$ & \small $2.0449$  \\
$\sigma=6$ & \small $1.4063$& \small$1.5515$ & \small $1.6345$ &
\small $1.6960$ & \small $1.7457$ &\small $1.7877$ & \small $1.8243$ &
\small $1.8569$  \\
\hline
\end{tabular}
\centerline{}   \caption{\small The exponential growth rates of
$k$-(nc), $\sigma$-(ca), RNA structures where $\sigma\ge 3$.
}\label{T:1}
\end{center}
\end{table}
To be concrete, the ratio of $2$-(nc) over $2$-crossing random
$3$-(nc) structures for $n=75$ is $\approx 6\times 10^{-5}$, while
the ratio of $2$-(nc) versus $2$-(nc) RNA structures generated by
{\sf cross} is $\approx 1.7027$. In other words, {\sf cross}
overrepresents $2$-(nc) structures at a rate of approximately $300\,
000\, :\, 1$. In Figure~\ref{F:33-34-data} we illustrate the fact
that $3$-(nc), $3$-(ca) mfe structures are more similar to $2$-(nc)
than to $3$-(nc) random structures.

%%%%
%%%%%%%%%%%%%%%%%%%%%%%%%%%%%%%%%%%%%%%%%%%%%%%%%%%%%%%%%%%%%%%%%%%%%%%%%%%%%%
%%%
%%%%
%%%%%%%%%%%%%%%%%%%%%%%%%%%%%%%%%%%%%%%%%%%%%%%%%%%%%%%%%%%%%%%%%%%%%%%%%%%%%%
%%%

%%%
%%%%%%%%%%%%%%%%%%%%%%%%%%%%%%%%%%%%%%%%%%%%%%%%%%%%%%%%%%%%%%%%%%%%
%%%%

\section{The distribution of returns}\label{S:return}

%%%
%%%%%%%%%%%%%%%%%%%%%%%%%%%%%%%%%%%%%%%%%%%%%%%%%%%%%%%%%%%%%%%%%%%%
%%%%

\begin{figure}[ht]
\centerline{%
\epsfig{file=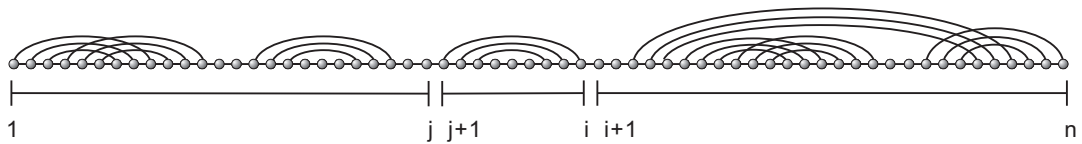,width=0.95\textwidth}\hskip15pt
 }
\caption{\small Three distinct segments: The first, $[1,j]$ and the
last, $[i+1,n]$ contains an arbitrary RNA structure, respectively.
While the second, $[j+1,i]$ contains an irreducible RNA structure.}
\label{F:hitgene}
\end{figure}

In this section we study the distribution of returns, i.e.~the endpoint
locations of irreducible substructures in RNA random and RNA mfe structures.
In other words, we compute the probability for a particular position to be
the {\it endpoint} of an irreducible substructure.
Let $\chi(s)$ denote the set of returns of a given structure $s$. Clearly,
for each return at $i$ there exists an irreducible substructure starting at
$j+1$ and ending at $i$.
Accordingly, a structure decomposes into $3$ distinct segments: the first
being an arbitrary substructure over the $[1,j]$,
the second being an irreducible substructure over $[j+1,i]$ and the third
being an arbitrary substructure over $[i+1,n]$, see Figure~\ref{F:hitgene}.
%%%
%%%%%%%%%%%%%%%%%%%%%%%%%%%%%%%%%%%%%%%%%%%%%%%%%%%%%%%%%%%%%%%%%%%%
%%%%
%%%
%%%%%%%%%%%%%%%%%%%%%%%%%%%%%%%%%%%%%%%%%%%%%%%%%%%%%%%%%%%%%%%%%%%%
%%%%
We denote the $n$-th coefficient of the generating function of ${\bf
R}_{k,\sigma}(z)$ by $R_{k,\sigma}(n)$, i.e., $R_{k,\sigma}(n)$ is
the number of irreducible $k$-(nc), $\sigma$-(ca) RNA structures
over length $n$. Let $T_{k,\sigma}(n)$ denote the coefficient ${\bf
T}_{k,\sigma}(z)$, i.e., $T_{k,\sigma}(n)$ is the number of
$k$-(nc), $\sigma$-(ca) RNA structures over length $n$. Finally, let
$a_i(n)$ denote the number of RNA structures of length $n$
containing $i$ as a return.
%%%
%%%%%%%%%%%%%%%%%%%%%%%%%%%%%%%%%%%%%%%%%%%%%%%%%%%%%%%%%%%%%%%%%%%%%%%%%%%%%%%
%%%
\begin{proposition}\label{P:return}
Let $k\ge 2$ and $\sigma\ge 1$ be natural numbers, $\mu=(k-1)^2+(k-1)/2$
and $\alpha_{k,\sigma}$ denote the unique dominant singularity of
${\bf T}_{k,\sigma}(z)$. Then
\begin{eqnarray*}
\mathbb{P}[i\in \chi(s)]&\sim & \left(1-\frac{i}{n}\right)^{-\mu}
\left[i^{-\mu}-(i-1)^{-\mu}\alpha_{k,\sigma}\right]
\end{eqnarray*}
and in particular for $i\to \infty$ and $(n-i)\to\infty$
\begin{eqnarray*}
\frac{\mathbb{P}[i+1\in\chi(s)]} {\mathbb{P}[i\in\chi(s)]} \sim
\left(\frac{n-i}{n-i-1}\right)^{\mu}.
\end{eqnarray*}
\end{proposition}
%%%
%%%%%%%%%%%%%%%%%%%%%%%%%%%%%%%%%%%%%%%%%%%%%%%%%%%%%%%%%%%%%%%%%%%%%%%%%%%%%%%
%%%

\begin{figure}[ht]
\centerline{%
\epsfig{file=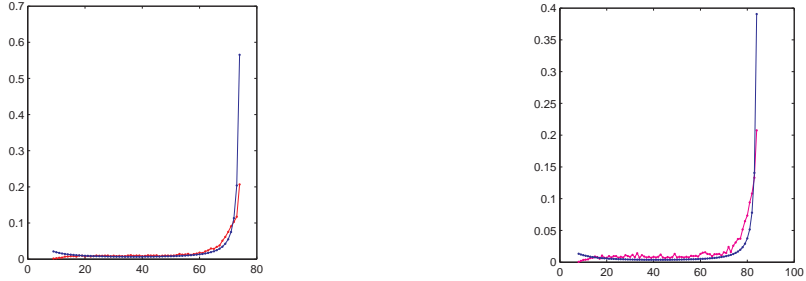,width=0.9\textwidth}\hskip15pt
 }
\caption{\small The distribution of returns for mfe and random
structures of finite size: the lhs shows $2$-(nc), $3$-(ca), mfe and
random structures of length $75$ (red/blue). The rhs displays $3$-(nc),
$3$-(ca), mfe and random structures of length $85$ (magenta/blue).
}\label{F:hitp12}
\end{figure}

Proposition~\ref{P:return} implies that returns are most likely to
occur at the end of the sequence, see Figure~\ref{F:hitp12}.
Furthermore the probability for the occurrence at the end of the
sequence is exponential with exponent $\mu=(k-1)^2+(k-1)/2$.
Consequently, larger crossing numbers imply that ``later'' returns
become more likely.

%%%
%%%%%%%%%%%%%%%%%%%%%%%%%%%%%%%%%%%%%%%%%%%%%%%%%%%%%%%%%%%%%%%%%%%%%%%%%%%%%
%%%
%%%
%%%%%%%%%%%%%%%%%%%%%%%%%%%%%%%%%%%%%%%%%%%%%%%%%%%%%%%%%%%%%%%%%%%%%%%%%%%%%
%%%

%%%
%%%%%%%%%%%%%%%%%%%%%%%%%%%%%%%%%%%%%%%%%%%%%%%%%%%%%%%%%%%%%%%%%%%%%%%%%
%%%

\section{Irreducible substructures}\label{S:sizecomp}

\begin{figure}[ht]
\centerline{%
\epsfig{file=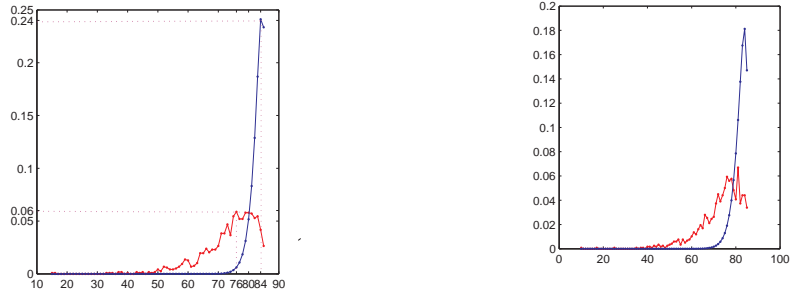,width=0.9\textwidth}\hskip15pt
 }
\caption{\small The distribution of the size of the unique
irreducible component for mfe and random structures of length $85$:
The lhs displays $2$-(nc), $3$-(ca) mfe (red) and random structures
(blue), the rhs shows $3$-(nc), $3$-(ca) mfe (red) and random
structures (blue).} \label{F:data2+3}
\end{figure}

\begin{figure}[ht]
\centerline{%
\epsfig{file=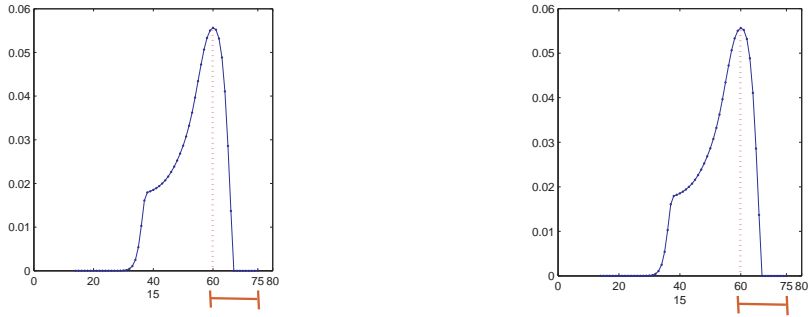,width=0.9\textwidth}\hskip15pt }
\caption{\small $(n-x_n)$ is independent $n$: we display the
distribution of the sizes of the giants conditional on the existence
of two irreducibles. The lhs shows $2$-(nc), $3$-(ca) random
structures for $n=75$ and the rhs for $n=85$, respectively. In both
distributions we highlight the distance (red) between the typical
size of the giant and the end of the sequence. } \label{F:sizeth4}
\end{figure}

%%%
%%%%%%%%%%%%%%%%%%%%%%%%%%%%%%%%%%%%%%%%%%%%%%%%%%%%%%%%%%%%%%%%%%%%%%%%%
%%%
In this section we compute the distribution of sizes of large
irreducible substructures. In Section~\ref{S:return} we established
that an irreducible substructure is typically ``large'', in the
following we shall prove substantial improvements: the size of the
largest, irreducible substructure is of size at least $n-O(1)$.
Section~\ref{S:comp1} and Section~\ref{S:return} show, that RNA
structures typically decompose into either one or two irreducible
components. We therefore restrict our analysis to these two
scenarios. We shall begin by studying the size, $x_n$, of an unique
irreducible RNA substructure.
%%%
%%%%%%%%%%%%%%%%%%%%%%%%%%%%%%%%%%%%%%%%%%%%%%%%%%%%%%%%%%%%%%%%%%%%%%%%%
%%%
\begin{lemma}\label{L:prob}
Suppose an RNA structure contains an unique, irreducible substructure, $s$,
then
\begin{eqnarray}
\mathbb{P}(\vert s\vert = x_n \mid \text{\rm $s$ is unique})\sim
\frac{(n-x_n+1)c\cdot
x_n^{-\mu}\alpha_{k,\sigma}^{-x_n}}{c\cdot\left(\frac{1}
{1-\alpha_{k,\sigma}}\right)^2n^{-\mu}\alpha_{k,\sigma}^{-n}}.
\end{eqnarray}
In particular, any unique irreducible substructure has size of at least
$n-O(1)$.
\end{lemma}
%%%
%%%%%%%%%%%%%%%%%%%%%%%%%%%%%%%%%%%%%%%%%%%%%%%%%%%%%%%%%%%%%%%%%%%%%%%%%
%%%
Furthermore for
\begin{equation}
\frac{1}{\alpha_{k,\sigma}^{-\frac{1}{\mu}}-1}\le
x_n\le n-\frac{\alpha_{k,\sigma}}{1-\alpha_{k,\sigma}}
\end{equation}
the conditional probability of having an unique irreducible substructure $s$
of size $x_n$ is strictly monotone in $x_n$ and for $x_n=n-\frac{\alpha_{k,
\sigma}}{1-\alpha_{k,\sigma}}$ given by
\begin{equation}\label{E:peak}
\lim_{n\rightarrow\infty}
\mathbb{P}\left(\vert s\vert= n-\frac{\alpha_{k,\sigma}}{1-\alpha_{k,\sigma}}
\mid\text{\rm $s$ is unique}\right)=
\left(1-\alpha_{k,\sigma}\right)\alpha_{k,\sigma}
^{\frac{\alpha_{k,\sigma}}{1-\alpha_{k,\sigma}}}.
\end{equation}
Lemma~\ref{L:prob} and the above observations show that a unique
largest irreducible component is of size almost $n$. The few,
remaining unpaired nucleotides are size $O(1)$, see
Figure~\ref{F:HCV}. The random structure distributions of length
$85$ derived in Lemma~\ref{L:prob} are given in
Figure~\ref{F:data2+3} together with the distributions for $2$- and
$3$-(nc), $3$-(ca) RNA mfe structures of length $n=85$. As for
random structures, we observe that in this case the dominant
singularities are $\alpha_{2,3}=0.6053$ and $\alpha_{3,3}=0.4914$
and the maximal probabilities as specified in eq.~(\ref{E:peak}) are
given by
\begin{eqnarray*}
\left(1-\alpha_{2,3}\right) \alpha_{2,3}^{\frac{\alpha_{2,3}}
{1-\alpha_{2,3}}} & \approx & 0.18\\
\left(1-\alpha_{3,3}\right)\alpha_{3,3}^{\frac{\alpha_{3,3}}
{1-\alpha_{3,3}}} & \approx & 0.25,
\end{eqnarray*}
respectively. The distributions of $2$- and $3$-(nc), $3$-(ca) mfe
structures exhibit similar features as those of random structures.
Since the data on mfe structures are obtained by sampling random
sequences having a unique, irreducible substructure, they represent
a refinement of the data given in Figure~\ref{F:hitp12}.
%%%
%%%%%%%%%%%%%%%%%%%%%%%%%%%%%%%%%%%%%%%%%%%%%%%%%%%%%%%%%%%%%%%%%%%%%%%%%%%%%%
%%%
%%%
%%%%%%%%%%%%%%%%%%%%%%%%%%%%%%%%%%%%%%%%%%%%%%%%%%%%%%%%%%%%%%%%%%%%%%%%%%%%%%
%%%
Next we consider the case of two irreducible substructures. As in the case of
a unique irreducible substructure, here, the larger of the two irreducibles
contains almost all nucleotides. The proofs, however, become
substantially more involved.
%%%
%%%%%%%%%%%%%%%%%%%%%%%%%%%%%%%%%%%%%%%%%%%%%%%%%%%%%%%%%%%%%%%%%%%%%%%%%%%%%%
%%%
\begin{figure}[ht]
\centerline{%
\epsfig{file=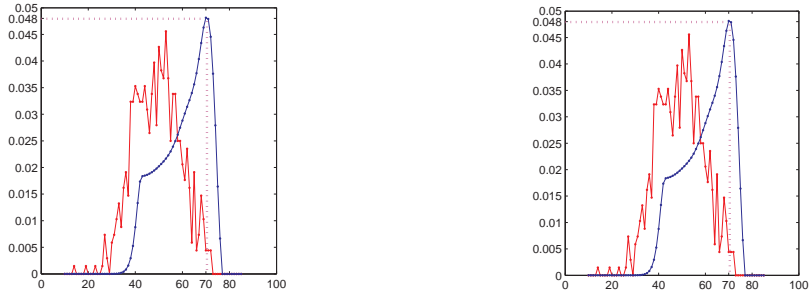,width=0.9\textwidth}\hskip15pt
 }
\caption{\small The distribution of the sizes of the giant for mfe and
random structures having exactly $2$ irreducible
components. The lhs: $2$-(nc), $3$-(ca) mfe structures (red) and
random structures for $n=85$ and $100$ (blue/black). The rhs:
$3$-(nc), $3$-(ca) mfe structures (red) and random structures for
$n=85$ and $100$ (blue/black).} \label{F:sizecom2}
\end{figure}
%%%
%%%%%%%%%%%%%%%%%%%%%%%%%%%%%%%%%%%%%%%%%%%%%%%%%%%%%%%%%%%%%%%%%%%%%%%%%%%%%%
%%%
%%%
%%%%%%%%%%%%%%%%%%%%%%%%%%%%%%%%%%%%%%%%%%%%%%%%%%%%%%%%%%%%%%%%%%%%%%%%%%%%%%
%%%
\begin{lemma}\label{L:2}
Suppose we are given an RNA structure $S$, that contains exactly the two
irreducible substructures, $s_1$ and $s_2$, where $x_n=\vert s_1\vert\ge
\vert s_2\vert$. Then
\begin{equation}
\mathbb{P}(\vert s_1\vert = x_n \mid \text{\rm $S$ contains $s_1,s_2$})\sim
\begin{cases}
o(1) &
 \text{\it for } (n-x_n)\rightarrow \infty\\
2(1-\alpha_{k,\sigma})^3\cdot c_{a}\cdot(\alpha_{k,\sigma})^{a}
&\text{\it for } (n-x_n)\rightarrow a<\infty,
\end{cases}
\end{equation}
where $\mu=(k-1)^2+(k-1)/2$, $c_a>0$ and $a\ge 1$. In particular, in
the limit of long sequences, $s_1$ has a.s.~a size of at least
$(n-O(1))$.
\end{lemma}
%%%
%%%%%%%%%%%%%%%%%%%%%%%%%%%%%%%%%%%%%%%%%%%%%%%%%%%%%%%%%%%%%%%%%%%%%%%%%%%%%%
%%%
According to Lemma~\ref{L:2}, we have for any $(n-x_n)\rightarrow
a<\infty$
\begin{eqnarray}
\lim_{n\rightarrow\infty} \mathbb{P}(\vert s_1\vert = x_n \mid
\text{\rm $S$ contains $s_1,s_2$}) &=& 2(1-\alpha_{k,\sigma})^3\cdot
c_{a}\cdot (\alpha_{k,\sigma})^{a}.
\end{eqnarray}
It follows from the proof of Lemma~\ref{L:2} in
Section~\ref{S:proof} that $ c_a=\sum_{i=1}^a{a-i+2\choose
2}R_{k,\sigma}(i) $. Let $\hat{a}$ be a positive constant for which
$c_{\hat{a}}\cdot (\alpha_{k,\sigma})^{\hat{a}}$ maximal.
Consequently the probability $\mathbb{P}(\vert s_1\vert = x_n \mid
\text{\rm $S$ contains $s_1,s_2$})$ is maximal at $x_n=n-\hat{a}$,
implying that the size of the largest irreducible component is in
the limit of long sequences typically $n-\hat{a}$, with probability
$(1-\alpha_{k,\sigma})^3 c_{\hat{a}}\cdot
(\alpha_{k,\sigma})^{\hat{a}}$. Note that for $3$-(nc), $3$-(ca)
random structures of length $n$, we have $\alpha_{3,3}=0.4914$.
Maximizing the term $c_{\hat{a}}\cdot (\alpha_{k,\sigma})^{\hat{a}}$
yields $\hat{a}=14$ and accordingly the size of the largest
component is likely to be $n-14$. Figure~\ref{F:sizecom2} confirms
that already for $n=85$, the size of the largest irreducible
component is typically $70$. Figure~\ref{F:sizecom2} shows first
that the probability $\mathbb{P}(\vert s_1\vert = x_n \mid \text{\rm
$S$ contains $s_1,s_2$})$ is sharply concentrated at
$x_n=n-\hat{a}$, as implied by Lemma~\ref{L:2}. Second, as $n$
increases, the distribution of component sizes shifts into a limit
distribution which is sharply concentrated at $n-\hat{a}$.
Furthermore we remark that, by construction, $\hat{a}$ is
independent of $n$, see Figure~\ref{F:sizeth4}. For $n=75$ and
$n=85$, the size of the largest irreducible component is localized
at $x_{75}=60$ and $x_{85}=70$.
%%%%%
%%%%%%%%%%%%%%%%%%%%%%%%%%%%%%%%
%%%%%
%%%%%
%%%%%%%%%%%%%%%%%%%%%%%%%%%%%%%%%%%%%%%%%%%%%%%%%%%%%%%%%%%%%%%%%%%%%%%%%%%%%%%
%%%%%
%%%%%
%%%%%%%%%%%%%%%%%%%%%%%%%%%%%%%%%%%%%%%%%%%%%%%%%%%%%%%%%%%%%%%%%%%%%%%%%%%%%%%
%%%%%
\section{Discussion}
%%%
%%%%%%%%%%%%%%%%%%%%%%%%%%%%%%%%%%%%%%%%%%%%%%%%%%%%%%%%%%%%%%%%%%%%%%%%%%%%%%%
%%%

Let us integrate our results and put them into context. Employing
the Motzkin-path (Figure~\ref{F:sec}) interpretation, it is
straightforward to construct random RNA secondary structures.
However, random RNA pseudoknot structures are a different matter.
Their inherent cross-serial dependencies
(Figure~\ref{F:plasmid.eps}) prohibit constructive recurrence
relations and despite their $D$-finiteness \cite{Reidys:07pseu}, at
present time, there exists no computer algorithm that can construct
a random RNA pseudoknot structure in polynomial time with uniform
probability. Consequently, any data on limit distributions of random
RNA pseudoknot structures are nontrivial and virtually impossible to
obtain computationally. In this paper we have shown that random RNA
structures decompose into a small number of irreducible
substructures, see Figures~\ref{F:plasmid.eps}, \ref{F:pseudo1} and
\ref{F:HCV}. We established in Section~\ref{S:sizecomp} that one of
these irreducibles is in fact a ``giant'', i.e.~it contains almost
all nucleotides. Key structural parameters, like the maximum number
of mutually crossing bonds, as well as the minimum stack size do not
fundamentally change this picture, see Figure~\ref{F:experiment} and
Figure~\ref{F:match}. In Section~\ref{S:comp1} we discussed the
distribution of random structures and mfe structures. We actually
used the limit of long sequences in order to prove a shift-result,
allowing for the reduction of $k$-(nc) $\sigma$-(ca) structures over
$n$ to $k$-(nc), $(\sigma-j)$-(ca) structures over $n-f(j)$.
Figure~\ref{F:shift} illustrates that the original and the
correspondingly shifted distributions of mfe structures virtually
``coincide''. Mfe and random structures exhibit significant
differences. Most striking maybe is the vast preference of
noncrossing RNA pseudoknot structures over their crossing
counterparts. While the percentage of $63\%$ of folded noncrossing
configurations for $n=75$ does not seem to be particularly
remarkable, Theorem~\ref{T:centra2} of Section~\ref{S:theory} shows,
that the above percentage is equivalent to a factor of
$300\,000\,:\,1$, relative to random sampling. This is certainly a
consequence of the currently implemented pseudoknot-loop energy
parameters. At this point it is pure speculation whether or not
different energy parameters or significantly longer sequence length
will alter this picture. In \cite{Reidys:bmc} the reader can find
more data on the fraction of noncrossing configurations of $3$-(nc)
mfe structures. In any case, the overrepresentation of noncrossing
configurations implies, that the distributions of $3$-(nc) mfe
structures are in fact more similar to those of random RNA secondary
structures.

Having established that, even in the limit of long sequences, only a
few irreducibles exist, the next question is to determine their
respective sizes. In Section~\ref{S:return} and
Section~\ref{S:sizecomp} we achieve this by studying returns, that
is the endpoints of irreducible substructures and the distribution
of their sizes in Lemma~\ref{L:prob} and Lemma~\ref{L:2}. For random
structures we observe, that the largest irreducible component is a
giant as it contains almost all nucleotides. For mfe structures we
observe a systematic shift towards smaller sizes of the giant,
however a giant irreducible substructure typically also exists in
mfe structures. Aside from these structural results we present in
Section~\ref{S:sizecomp}, eq.~(\ref{E:peak}), a simple formula for
identifying the typical size of an unique giant and confirm in
Figure~\ref{F:data2+3} its applicability to mfe structures. Along
these lines we furthermore localize the typical size of the giant in
case of two irreducible substructures. In addition we make its
dependence on $k$, $\sigma$ and $n$ explicit.

%%%
%%%%%%%%%%%%%%%%%%%%%%%%%%%%%%%%%%%%%%%%%%%%%%%%%%%%%%%%%%%%%%%%%%%%%%%%%%%%%%%
%%%

\section{Proofs}\label{S:proof}

%%%
%%%%%%%%%%%%%%%%%%%%%%%%%%%%%%%%%%%%%%%%%%%%%%%%%%%%%%%%%%%%%%%%%%%%%%%%%%%%%%%
%%%
{\bf Proof of Proposition~\ref{P:elong}.}
\begin{proof}
$D$-finiteness \cite{Stanley:80} of ${\bf R}_{k,\sigma}(z)$ guarantees the
existence of analytic continuation in some simply connected domain containing
zero around the dominant singularity $\alpha_{k,\sigma}$. Therefore the
singular expansion of ${\bf R}_{k,\sigma}(z)$ at its dominant singularity
$\alpha_{k,\sigma}$ exists and in case of $k\equiv1\mod2$ we have, setting
$\mu=(k-1)^2+\frac{k-1}{2}$,
\begin{eqnarray}\label{E:odd}
{\bf R}_{k,\sigma}(z)=
\tau_k-c_k\left(1-\frac{z}{\alpha_{k,\sigma}}\right)^{\mu-1}
\ln\left(1-\frac{z}{\alpha_{k,\sigma}}\right)(1+o(1)), \quad &
\text{\rm where} \ c_k >0.
\end{eqnarray}
In case of $k\equiv0\mod2$, the singular expansion is given by
\begin{eqnarray}\label{E:even}
{\bf R}_{k,\sigma}(z)=
\tau_k-c_k\left(1-\frac{z}{\alpha_{k,\sigma}}\right)^{\mu-1}(1+o(1)),
\quad & \text{\rm where} \ c_k >0.
\end{eqnarray}
In the following we restrict ourselves to the analysis of the case
$k\equiv 1\mod2$. The arguments for $k\equiv 0\mod 2$ are completely
analogous. From the singular expansion of ${\bf R}_{k,\sigma}(z)$ we
derive
\begin{eqnarray*}
R_{k,\sigma}(n)\sim
n^{-\mu}\left(\frac{1}{\alpha_{k,\sigma}}\right)^n
\end{eqnarray*}
and, apparently, $R_{k,\sigma+1}(n)<R_{k,\sigma}(n)$. For sufficiently large
$n$, we have
\begin{eqnarray*}
\frac{R_{k,\sigma+1}(n+x_{n}^{(k)})}{R_{k,\sigma}(n)} & = &
\frac{(n+x_{n}^{(k)})^{-\mu}\left(\frac{1}{\alpha_{k,\sigma+1}}\right)^{n+x_{n}^{(k)}}}
{n^{-\mu}\left(\frac{1}{\alpha_{k,\sigma}}\right)^n}\\
& = &
\left(1+\frac{x_{n}^{(k)}}{n}\right)^{-\mu}\cdot\frac{(\alpha_{k,\sigma})^n}
{(\alpha_{k,\sigma+1})^{n+x_{n}^{(k)}}}.
\end{eqnarray*}
Suppose $\lim_{n\to\infty}(\frac{R_{k,\sigma+1}(n+x_{n}^{(k)})}
{R_{k,\sigma}(n)}) = c>0$, then
\begin{equation}\label{E:gut}
x_{n}^{(k)}=\left\lfloor n\left(
\frac{\ln(\alpha_{k,\sigma})}{\ln(\alpha_{k,\sigma+1})}-1\right)\right\rfloor.
\end{equation}
Accordingly
\begin{eqnarray}
\frac{R_{k,\sigma+1}(n+x_{n}^{(k)})}{R_{k,\sigma}(n)}\sim
\left(\frac{\ln(\alpha_{k,\sigma})}
{\ln(\alpha_{k,\sigma+1})}\right)^{-\mu}<1
\end{eqnarray}
and eq.~(\ref{E:elong}) follows. Using the singular expansion of ${\bf
R}_{k,\sigma}(z)$ given in eq.~(\ref{E:odd}) we derive
\begin{eqnarray*}
\beta_{k,\sigma,j}(n)&=&[z^n]{\bf
R}_{k,\sigma}(z)^j\left(\frac{1}{1-z}\right)^{j+1}\\
&=&[z^n]\left[\tau_k-c_k\left(1-\frac{z}{\alpha_{k,\sigma}}\right)^{\mu-1}
\ln\left(1-\frac{z}{\alpha_{k,\sigma}}\right)(1+o(1))\right]^{j}
\left(\frac{1}{1-z}\right)^{j+1}\\
&\sim& c_{k,\sigma}^{(1)}\cdot n^{-\mu}\alpha_{k,\sigma}^{-n} \qquad
\mbox{ for some constant } c_{k,\sigma}^{(1)}>0
\end{eqnarray*}
and consequently
\begin{eqnarray*}
\beta_{k,\sigma+1,j}(n+y_{n,k}) &\sim& c_{k,\sigma+1}^{(1)}\cdot
(n+y_{n,k})^{-\mu}\alpha_{k,\sigma+1}^{-n-y_{n,k}} \qquad \mbox{ for
some constant } c_{k,\sigma+1}^{(1)}>0.
\end{eqnarray*}
We observe that
$\beta_{k,\sigma,j}(n)\approx\beta_{k,\sigma+1,j}(n+y_{n,k})$ holds
only if $y_{n,k}=\lfloor c_{k}\cdot n\rfloor$ for some constant
$c_{k}>0$ with the property
\begin{eqnarray}{\label{E:scalefac}}
\left(\frac{\alpha_{k,\sigma}}{\left(\alpha_{k,\sigma+1}\right)
^{1+c_{k}}}\right)^n
=\frac{c_{k,\sigma}^{(1)}}{c_{k,\sigma+1}^{(1)}}(1+c_k)^{\mu}.
\end{eqnarray}
The solution $c_k$ of eq.~(\ref{E:scalefac}) is asymptotically
\begin{equation*}
c_{k}\approx\frac{\ln(\alpha_{k,\sigma})}
{\ln(\alpha_{k,\sigma+1})},
\end{equation*}
whence $y_{n,k}=x_{n}^{(k)}$ and eq.~(\ref{E:elongirre}) is
established.
\end{proof}

{\bf Proof of Proposition~\ref{P:return}.}
\begin{proof}
We begin by splitting a given structure $s$ at $j$ and $i$:
\begin{eqnarray*}
a_i=\sum_{j\le i}T_{k,\sigma}(j)\cdot R_{k,\sigma}(i-j)\cdot
T_{k,\sigma}(n-i)&=&T_{k,\sigma}(n-i)\cdot
\sum_{j\le i}T_{k,\sigma}(j)\cdot R_{k,\sigma}(i-j)\\
\end{eqnarray*}
and $\sum_{j\le i}T_{k,\sigma}(j)\cdot
R_{k,\sigma}(i-j)=[z^i]\left({\bf T}_{k,\sigma}(z){\bf
R}_{k,\sigma}(z)\right)$. Furthermore, we derive
\begin{eqnarray*}
\sum_{i=1}^n a_i&=&\sum_{i=1}^n T_{k,\sigma}(n-i)\cdot
[z^i]\left({\bf
T}_{k,\sigma}(z){\bf R}_{k,\sigma}(z)\right)\\
&=&\sum_{i=1}^n [z^{n-i}]{\bf T}_{k,\sigma}(z)\cdot [z^i]\left({\bf
T}_{k,\sigma}(z){\bf R}_{k,\sigma}(z)\right)\\
&=&[z^n]{\bf T}_{k,\sigma}^2(z){\bf R}_{k,\sigma}(z).
\end{eqnarray*}
Consequently, the probability of a return at position $i$ is given by
\begin{eqnarray}
\mathbb{P}[i\in\chi(s)]&=&\frac{T_{k,\sigma}(n-i)\cdot
[z^i]\left({\bf T}_{k,\sigma}(z){\bf
R}_{k,\sigma}(z)\right)}{[z^n]{\bf T}_{k,\sigma}^2(z){\bf
R}_{k,\sigma}(z)}.
\end{eqnarray}
Using ${\bf R}_{k,\sigma}(z)=1-z-\frac{1}{{\bf T}_{k,\sigma}(z)}$ we
rewrite the probability $\mathbb{P}[i\in\chi(s)]$ as
\begin{eqnarray}
\mathbb{P}[i\in\chi(s)]&=&\frac{T_{k,\sigma}(n-i)\cdot
[z^i]\left((1-z){\bf T}_{k,\sigma}(z)-1\right)}{[z^n](1-z){\bf
T}_{k,\sigma}^2(z)-{\bf T}_{k,\sigma}(z)}.
\end{eqnarray}
We next use the singular expansion of ${\bf T}_{k,\sigma}(z)$ at
the dominant singularity $\alpha_{k,\sigma}$
\begin{eqnarray}\label{E:asy}
{\bf T}_{k,\sigma}(z)=
\begin{cases}
O((1-\frac{z}{\alpha_{k,\sigma}})^{\mu-1} \ln(1-\frac{z}{\alpha_{k,\sigma}})) &
\text{\rm for $k$ odd as
$z\rightarrow \alpha_{k,\sigma}$} \\
O((1-\frac{z}{\alpha_{k,\sigma}})^{\mu-1})             & \text{\rm for $k$
 even as $z\rightarrow \alpha_{k,\sigma}$,}
\end{cases}\label{E:W_k}
\end{eqnarray}
where $\mu=(k-1)^2+(k-1)/2$. We restrict ourselves proving the case of
$k\equiv 1\mod 2$, the case $k\equiv 0\mod 2$ follows analogously.
Since $D$-finite power series
form an algebra,
${\bf Q}_{k,\sigma}(z)=(1-z){\bf T}_{k,\sigma}^2(z)-{\bf T}_{k,\sigma}(z)$ is
$D$-finite and analytic continuation and singular expansion exist.
The latter is given by
\begin{align*}
{\bf Q}_{k,\sigma}(z)
&=(1-z)O\left(\left[(1-\frac{z}{\alpha_k})^{\mu-1}
\ln(1-\frac{z}{\alpha_k})\right]^2\right)-O\left((1-\frac{z}{\alpha_k})^{\mu-1}
\ln(1-\frac{z}{\alpha_k})\right), \quad z\rightarrow \alpha_{k,\sigma} \\
&=O\left((1-\frac{z}{\alpha_k})^{\mu-1}
\ln(1-\frac{z}{\alpha_k})\right), \quad z\rightarrow \alpha_{k,\sigma}.
\end{align*}
Therefore, extracting the coefficients of the singular expansion, we obtain
\begin{equation}
[z^n]{\bf Q}_{k,\sigma}(z)\sim
n^{-\mu}\alpha_{k,\sigma}^{-n}.
\end{equation}
Using $[z^i]{\bf T}_{k,\sigma}(z)=T(i)\sim i^{-\mu}\alpha_{k,\sigma}^{-i}$ as
$i\rightarrow \infty$ we arrive at
\begin{eqnarray*}
\mathbb{P}[i\in\chi(s)]&\sim& \frac{T(n-i)\left[[z^i]{\bf
T}_{k,\sigma}(z)-[z^{i-1}]
{\bf T}_{k,\sigma}(z)\right]}{[z^n]{\bf Q}_{k,\sigma}(z)}\\
&\sim &
\frac{(n-i)^{-\mu}\alpha_{k,\sigma}^{-n+i}[i^{-\mu}\alpha_{k,\sigma}^{-i}
-(i-1)^{-\mu}\alpha_{k,\sigma}^{-i+1}]}
{n^{-\mu}\alpha_{k,\sigma}^{-n}}\\
&=&\left(1-\frac{i}{n}\right)^{-\mu}\left[i^{-\mu}-(i-1)^{-\mu}
\alpha_{k,\sigma}\right].
\end{eqnarray*}
From this we immediately conclude in case of $i\to \infty$ and
$(n-i)\to\infty$
\begin{eqnarray*}
\frac{\mathbb{P}[i+1\in\chi(s)]}{\mathbb{P}[i\in\chi(s)]} &\sim
&\frac{\left(1-\frac{i+1}{n}\right)^{-\mu}
\left[(i+1)^{-\mu}-i^{-\mu}\alpha_{k,\sigma}\right]}
{\left(1-\frac{i}{n}\right)^{-\mu}
\left[i^{-\mu}-(i-1)^{-\mu}\alpha_{k,\sigma}\right]}\\
&\sim &\left(\frac{n-i}{n-i-1}\right)^{\mu}\qquad i\to \infty,(n-i)\to\infty .
\end{eqnarray*}
\end{proof}

{\bf Proof of Lemma~\ref{L:prob}.}
\begin{proof}
Using the singular expansion of ${\bf R}_{k,\sigma}(z)$, eq.~(\ref{E:odd})
and eq.~(\ref{E:even}) we obtain
\begin{equation}\label{E:geneden}
\delta_{n,1} = [z^n]\left(\frac{1}{1-z}\right)^2{\bf R}_{k,\sigma}(z)
\sim c\cdot\left(\frac{1}{1-\alpha_{k,\sigma}}\right)^2
n^{-\mu}\alpha_{k,\sigma}^{-n}, \quad \mbox{ for some } c>0.
\end{equation}
We proceed by computing
\begin{equation}\label{E:genenom}
\delta_{n,1,x_n} = [z^{x_n}]{\bf R}_{k,\sigma}(z)\cdot
[z^{n-x_n}]\left(\frac{1}{1-z}\right)^2
\sim (n-x_n+1)c\cdot x_n^{-\mu}\alpha_{k,\sigma}^{-x_n}
\end{equation}
and combining eq.~(\ref{E:geneden}) and eq.~(\ref{E:genenom}), we arrive at
\begin{equation}\label{E:inspect}
\frac{\delta_{n,1,x_n}}{\delta_{n,1}}\sim
\frac{(n-x_n+1)c\cdot x_n^{-\mu}\alpha_{k,\sigma}^{-x_n}}
{c\cdot\left(\frac{1}{1-\alpha_{k,\sigma}}\right)^2
n^{-\mu}\alpha_{k,\sigma}^{-n}}.
\end{equation}
The critical term here in eq.~(\ref{E:inspect}) is readily identified to be
$\alpha_{k,\sigma}^{n-x_n}$ and consequently
\begin{equation}
\lim_{n\to\infty}\frac{\delta_{n,1,x_n}}{\delta_{n,1}}>0\quad\Longrightarrow
\quad x_n=n-O(1),
\end{equation}
whence the lemma.
\end{proof}

{\bf Proof of Lemma~\ref{L:2}.}
\begin{proof}
We distinguish the cases $x_n<\frac{n}{2}$ and $x_n\ge\frac{n}{2}$.
In case of $x_n<\frac{n}{2}$ we have
\begin{equation}
\delta_{n,2} = [z^n]\left(\frac{1}{1-z}\right)^3{\bf R}_{k,\sigma}^2(z)
\sim c\cdot\left(\frac{1}{1-\alpha_{k,\sigma}}\right)^3
n^{-\mu}\alpha_{k,\sigma}^{-n}, \quad \mbox{ for some } c>0.
\end{equation}
Therefore the number of structures containing $s_1$ of size $x_n$ is given by
\begin{eqnarray*}
\delta_{n,2,x_n}&=&\sum_{\max(s,t)=x_n}[z^{s}]{\bf
R}_{k,\sigma}(z)\cdot[z^{t}]{\bf R}_{k,\sigma}(z)\cdot
[z^{n-s-t}]\left(\frac{1}{1-z}\right)^3\\
&=&2[z^{x_n}]{\bf R}_{k,\sigma}(z)\cdot \sum_{t=1}^{x_n}[z^t]{\bf
R}_{k,\sigma}(z)\cdot[z^{n-x_n-t}]\left(\frac{1}{1-z}\right)^3
\end{eqnarray*}
The term $[z^{n-x_n-t}]\left(\frac{1}{1-z}\right)^3$ represents the number of
compositions of the integer $u=n-x_n-t$ into at most $3$ distinct parts,
denoted by $P(u,3)$. Assuming the first part to be $i$, ranging from
$1$ to $u$, the number of ways of dividing $(u-i)$ into at most
$2$ parts is $(u-i+1)$, whence
\begin{equation*}
[z^{u}]\left(\frac{1}{1-z}\right)^3=\sum_{i=0}^{u}P(u-i,2)=
\sum_{i=0}^{u}(u-i+1)={u+2 \choose 2}.
\end{equation*}
Consequently, we can rewrite $\delta_{n,2,x_n}$ as
\begin{equation}
\delta_{n,2,x_n} = 2[z^{x_n}]{\bf R}_{k,\sigma}(z)\cdot
\sum_{t=1}^{x_n}{n-x_n+2\choose 2} \cdot[z^t]{\bf R}_{k,\sigma}(z).
\end{equation}
{\it Claim.} Suppose $x_n$ and $n-x_n$ tend to infinity, as $n$ tends to
infinity. Then there exists some constant $\kappa>0$ such
that
\begin{eqnarray}
\sum_{i=1}^{x_n}{n-x_n-i+2\choose 2}[z^i]{\bf
R}_{k,\sigma}(z)=\kappa\cdot{n-2x_n+2\choose 2}[z^{x_n}]{\bf
R}_{k,\sigma}(z).
\end{eqnarray}
According to the Claim
\begin{eqnarray*}
\delta_{n,2,x_n}&=&2[z^{x_n}]{\bf R}_{k,\sigma}(z)\cdot
\sum_{t=1}^{x_n}{n-x_n-t+2\choose 2}
\cdot[z^t]{\bf R}_{k,\sigma}(z)\\
&\sim& 2[z^{x_n}]{\bf R}_{k,\sigma}(z)\cdot\kappa{n-2x_n+2\choose 2}
[z^{x_n}]{\bf
R}_{k,\sigma}(z)\\
&\sim& 2\kappa c^2 \cdot
x_n^{-2\mu}\alpha_{k,\sigma}^{-2x_n}{n-2x_n+2\choose 2},
\end{eqnarray*}
whence the probability of containing the largest component of size
$x_n<\frac{n}{2}$ is given by
\begin{equation}
\frac{\delta_{n,2,x_n}}{\delta_{n,2}}\sim \frac{2\kappa c^2 \cdot
x_n^{-2\mu}\alpha_{k,\sigma}^{-2x_n}{n-2x_n+2\choose
2}}{c\cdot\left(\frac{1}{1-\alpha_{k,\sigma}}\right)^3
n^{-\mu}\alpha_{k,\sigma}^{-n}}=o(1).
\end{equation}
In case of $x_n\ge \frac{n}{2}$ we derive,
\begin{eqnarray*}
\delta_{n,2,x_n}&=&\sum_{\max(s,t)=x_n}[z^{s}]{\bf
R}_{k,\sigma}(z)\cdot[z^{t}]{\bf R}_{k,\sigma}(z)\cdot
[z^{n-s-t}]\left(\frac{1}{1-z}\right)^3\\
&=&2[z^{x_n}]{\bf R}_{k,\sigma}(z)\cdot \sum_{t=1}^{n-x_n}[z^t]{\bf
R}_{k,\sigma}(z)\cdot[z^{n-x_n-t}]\left(\frac{1}{1-z}\right)^3\\
&=&2[z^{x_n}]{\bf R}_{k,\sigma}(z)\cdot[z^{n-x_n}]{\bf
R}_{k,\sigma}(z)\left(\frac{1}{1-z}\right)^3.
\end{eqnarray*}
Suppose $(n-x_n)\rightarrow\infty$, then the singular expansion implies
\begin{eqnarray*}
[z^{n-x_n}]{\bf R}_{k,\sigma}(z)\left(\frac{1}{1-z}\right)^3
&=&\kappa\cdot
\left(n-x_n\right)^{-\mu}\left(\alpha_{k,\sigma}\right)^{-n+x_n}.
\end{eqnarray*}
Accordingly, we derive the following asymptotic expression for
$\delta_{n,2,x_n}$
\begin{eqnarray*}
\delta_{n,2,x_n} &=& 2[z^{x_n}]{\bf
R}_{k,\sigma}(z)\cdot[z^{n-x_n}]{\bf
R}_{k,\sigma}(z)\left(\frac{1}{1-z}\right)^3\\
&\sim&  2c\kappa \cdot x_n^{-\mu}\alpha_{k,\sigma}^{-x_n}
(n-x_n)^{-\mu}\alpha_{k,\sigma}^{-(n-x_n)},
\end{eqnarray*}
Therefore we arrive at
\begin{equation}
\frac{\delta_{n,2,x_n}}{\delta_{n,2}}\sim \frac{2c\kappa \cdot
x_n^{-\mu}\alpha_{k,\sigma}^{-x_n}
(n-x_n)^{-\mu}\alpha_{k,\sigma}^{-(n-x_n)}}
{c\cdot\left(\frac{1}{1-\alpha_{k,\sigma}}\right)^3
n^{-\mu}\alpha_{k,\sigma}^{-n}} = 2\kappa (1-\alpha_{k,\sigma})^3
                        \left[\frac{x_n(n-x_n)}{n}\right]^{-\mu}.
\end{equation}
Note that $(n-x_n)\rightarrow\infty$ implies, that
$\lim_{n\rightarrow\infty}\frac{x_n}{n}=\nu$ for $\frac{1}{2}\le
\nu<1$. Consequently
\begin{equation}
\frac{\delta_{n,2,x_n}}{\delta_{n,2}}\sim 2\kappa
(1-\alpha_{k,\sigma})^3\left[\frac{x_n(n-x_n)}{n}\right]^{-\mu}
=2\kappa (1-\alpha_{k,\sigma})^3\left[\nu\cdot
n(1-\nu)\right]^{-\mu}=o(1),
\end{equation}
from which we immediately conclude
\begin{equation}
\lim_{n\to \infty}\frac{\delta_{n,2,x_n}}{\delta_{n,2}}=0
\quad\mbox{ for } x_n\ge\frac{n}{2} \mbox{ and }
(n-x_n)\rightarrow\infty.
\end{equation}
In case of $(n-x_n)\rightarrow a<\infty$, we set
\begin{equation}
c_a=\sum_{i=1}^a{a-i+2\choose 2}R_{k,\sigma}(i)=[z^a]{\bf
R}_{k,\sigma}(z)\left(\frac{1}{1-z}\right)^3.
\end{equation}
Accordingly
\begin{eqnarray*}
\delta_{n,2,x_n}&=&2[z^{x_n}]{\bf
R}_{k,\sigma}(z)\cdot[z^{n-x_n}]{\bf
R}_{k,\sigma}(z)\left(\frac{1}{1-z}\right)^3\\
&\sim&2c_a\cdot c \cdot
x_n^{-\mu}\left(\alpha_{k,\sigma}\right)^{-x_n}.
\end{eqnarray*}
We accordingly arrive at
\begin{equation}
\frac{\delta_{n,2,x_n}}{\delta_{n,2}}\sim\frac{2c_a\cdot c \cdot
x_n^{-\mu}\left(\alpha_{k,\sigma}\right)^{-x_n}}
{c\cdot\left(\frac{1}{1-\alpha_{k,\sigma}}\right)^3
n^{-\mu}\alpha_{k,\sigma}^{-n}}=2c_a\cdot(1-\alpha_{k,\sigma})^3\cdot
(\alpha_{k,\sigma})^{a},
\end{equation}
We observe that $x_n<\frac{n}{2}$ implies $(n-x_n)\rightarrow
\infty$, therefore we conclude in the case $(n-x_n)\rightarrow
\infty$,
$$\lim_{n\rightarrow\infty}\frac{\delta_{n,2,x_n}}{\delta_{n,2}}=0.$$
While $(n-x_n)\rightarrow a<\infty$ implies $x_n\ge\frac{n}{2}$, we
conclude that in the case $(n-x_n)\rightarrow a<\infty$,
\begin{equation*}
\lim_{n\rightarrow\infty}\frac{\delta_{n,2,x_n}}{\delta_{n,2}}
=2c_a\cdot(1-\alpha_{k,\sigma})^3\cdot (\alpha_{k,\sigma})^{a}.
\end{equation*}
completing the proof of the lemma.
\end{proof}

{\bf Proof of the Claim.}
\begin{proof}
Set
\begin{equation}
A_i={n-x_n-i+2\choose 2}[z^i]{\bf R}_{k,\sigma}(z)=
{n-x_n-i+2\choose 2}R_{k,\sigma}(i)
\end{equation}
We first show that $A_{x_n}$ is the maximal term $A_i$ for $1\le i\le x_n$.
In view of the fact that $x_n<\frac{n}{2}$,
\begin{eqnarray*}
\lim_{n\rightarrow \infty}\frac{A_{i+1}}{A_{i}}&=&\lim_{n\rightarrow
\infty}\frac{n-x_n-i}{n-x_n-i+2}\cdot
\frac{R_{k,\sigma}(i+1)}{R_{k,\sigma}(i)}\\
&=&\lim_{n\rightarrow
\infty}\frac{1}{1+\frac{2}{n-x_n-i}}
\frac{R_{k,\sigma}(i+1)}{R_{k,\sigma}(i)}\\
&>&\lim_{n\rightarrow\infty}\frac{1}{1+\frac{2}{\frac{n}{2}-i}}
\frac{R_{k,\sigma}(i+1)}{R_{k,\sigma}(i)}\\
&=&\frac{R_{k,\sigma}(i+1)}{R_{k,\sigma}(i)}>1.
\end{eqnarray*}
Therefore $A_{x_n}$ is maximal. We shall show next
\begin{eqnarray}{\label{E:sumsmall}}
\forall \; 0<\alpha<1;\qquad
\sum_{i<x_n-n^{\alpha}}A_i=o(1)\cdot A_{x_n}.
\end{eqnarray}
In view of $\sum_{i<x_n-n^{\alpha}}A_i<(x_n-n^{\alpha})A_{x_{n}-n^{\alpha}}$,
we obtain
\begin{equation*}
\frac{\sum_{i<x_n-n^{\alpha}}A_i}{A_{x_n}} < \frac{(x_n-n^{\alpha})
A_{x_{n}-n^{\alpha}}}{A_{x_n}}
=\frac{(x_n-n^{\alpha}){n-2x_n+n^{\alpha}+2 \choose
2}R_{k,\sigma}(x_n-n^{\alpha})} {{n-2x_n+2 \choose
2}R_{k,\sigma}(x_n)}
\end{equation*}
Using $R_{k,\sigma}(n)\sim c_1 n^{-\mu}\alpha_{k,\sigma}^{-n}$, for some
$c_1>0$, we arrive at
\begin{eqnarray*}
\frac{\sum_{i<x_n-n^{\alpha}}A_i}{A_{x_n}}&<&
\frac{(x_n-n^{\alpha}){n-2x_n+n^{\alpha}+2
\choose 2}R_{k,\sigma}(x_n-n^{\alpha})} {{n-2x_n+2 \choose
2}R_{k,\sigma}(x_n)}\\
&\sim&\left(1-\frac{n^{\alpha}}{x_n}\right)^{-\mu}
\left(x_n-n^{\alpha}\right)\left[1+\frac{n^{\alpha}}{n-2x_n+2}\right]
\left[1+\frac{n^{\alpha}}{n-2x_n+1}\right]\alpha_{k,\sigma}^{n^{\alpha}}=o(1),
\end{eqnarray*}
whence eq.~(\ref{E:sumsmall}). Next we claim that the terms close to $x_n$
contribute at most $O(A_{x_n})$, i.e.
\begin{eqnarray}
\sum_{j=0}^{n^{\alpha}}A_{x_n-j}=O(A_{x_n}).
\end{eqnarray}
To prove this, we compute for $1\le j\le n^{\alpha}$
\begin{eqnarray*}
\frac{A_{x_n-j}}{A_{x_n}}&\sim&\left[1+\frac{j}{n-2x_n+2}\right]\cdot
\left[1+\frac{j}{n-2x_n+1}\right]\cdot
\left[1-\frac{j}{x_n}\right]^{-\mu}\alpha_{k,\sigma}^{j}\\
&\le&\left[1+\frac{j}{n-2x_n+2}\right]\cdot
\left[1+\frac{j}{n-2x_n+1}\right]\alpha_{k,\sigma}^{j}\\
&<&\left(1+\frac{j}{2}\right)(1+j)\alpha_{k,\sigma}^j.
\end{eqnarray*}
Taking the sum over all $j$ we obtain
\begin{eqnarray*}
\sum_{j=0}^{n^{\alpha}}\frac{A_{x_n-j}}{A_{x_n}}&<&
\sum_{j=0}^{n^{\alpha}}\left(1+\frac{j}{2}\right)(1+j)\alpha_{k,\sigma}^j\\
&=&\frac{\alpha_{k,\sigma}^{n^{\alpha}+1}\left[(n^{\alpha}+1)^2
(\alpha_{k,\sigma}-1)^2(n^{\alpha}+1)(\alpha_{k,\sigma}-2)^2-
n^{\alpha}+1\right]-2}{2(\alpha_{k,\sigma}-1)^3},
\end{eqnarray*}
whence
\begin{eqnarray*}
\lim_{n\rightarrow \infty}\sum_{j=0}^{n^{\alpha}}\frac{A_{x_n-j}}{A_{x_n}}&<&
\frac{1}{(1-\alpha_{k,\sigma})^3}.
\end{eqnarray*}
Therefore, we obtain $\sum_{j=0}^{n^{\alpha}}A_{x_n-j}<
\frac{1}{(1-\alpha_{k,\sigma})^3}A_{x_n}$ and we arrive at
\begin{equation}
\sum_{i=0}^{x_n}A_i = \sum_{i<x_n-n^{\alpha}}A_i+\sum_{i\ge x_n-n^{\alpha}}A_i
= o(A_{x_n})+O(A_{x_n})=\kappa\cdot A_{x_n},
\end{equation}
for some constant $\kappa>0$, proving the Claim.
\end{proof}

%%%
%%%
%%%%%%%%%%%%%%%%%%%%%%%%%%%%%%%%%%%%%%%%%%%%%%%%%%%%%%%%%%%%%%%%%%%%%%%%
%%%
{\bf Acknowledgments.}
%%%
%%%%%%%%%%%%%%%%%%%%%%%%%%%%%%%%%%%%%%%%%%%%%%%%%%%%%%%%%%%%%%%%%%%%%%%%%%
%%%
We are grateful to W.Y.C.~Chen for stimulating discussions. Many
thanks to J.Z.M.~Gao for his help. This work was supported by the
973 Project, the PCSIRT Project of the Ministry of Education, the
Ministry of Science and Technology, and the National Science
Foundation of China.
%%%
%%%%%%%%%%%%%%%%%%%%%%%%%%%%%%%%%%%%%%%%%%%%%%%%%%%%%%%%%%%%%%%%%%%%%%%%
%%%

\end{document}